\documentclass[prd,showpacs,preprint]{revtex4}
\usepackage{graphicx,color}

\begin{document}

\title{Twist-3 distribution amplitudes of scalar mesons from QCD sum rules}

\author{Cai-Dian L\"{u} $^{1,2}$} \author{Yu-Ming Wang$^{2}$}
\author{Hao Zou$^{2}$}

\vspace*{1.0cm}

\affiliation{$^{1}$CCAST (World Laboratory), P.O. Box 8730, Beijing
100080, China;}

\affiliation{$^{2}$Institute of High Energy Physics, P.O. Box
918(4), Beijing 100049, China}

\vspace*{1.0cm}

\date{\today}
\begin{abstract}
We study the twist-3 distribution amplitudes  for scalar mesons made
up of two valence quarks based on QCD sum rules.
 By choosing the proper correlation functions, we derive the
moments of the scalar mesons up to the first two order.  Making use
of these moments, we then calculate the first two Gegenbauer
coefficients for twist-3 distribution amplitudes of scalar mesons.
It is found that the second Gegenbauer coefficients of scalar
density twist-3 distribution amplitudes for $K^{*}_0$ and $f_0$
mesons are quite close to that for $a_0$, which indicates that the
SU(3) symmetry breaking effect is tiny here. However, this effect
could not be neglected for the forth Gegenbauer coefficients of
scalar twist-3 distribution amplitudes between $a_0$ and $f_0$.
Besides, we also observe that the first two Gegenbauer coefficients
corresponding to the  tensor current twist-3 distribution amplitudes
for all the $a_0$, $K^{*}_0$ and $f_0$ are very small. The
renormalization group evolution of condensates, quark masses, decay
constants and moments are considered in our calculations. As a
byproduct, it is found that the masses for isospin I=1, ${1 \over
2}$ scalar mesons are around $1.27 \sim 1.41$ GeV and $1.44 \sim
1.56$ GeV respectively, while the mass for isospin state composed of
$\bar{s} s$ is $1.62 \sim 1.73$ GeV.

\end{abstract}

\pacs{11.55.Hx, 14.40.Cs} \maketitle


\section{Introduction}

Although the quark model has achieved great successes for several
decades,  the fundamental structures of scalar mesons are still
controversial. So far, there is not a definite answer on whether
they are two-quark states, multiquark states or even glueball,
molecule states among the light scalars yet \cite{scalar meson
1,scalar meson 2,scalar meson 3,scalar meson 4}. Much efforts have
been given to the study of decay and production of these mesons.
However, many theoretical predictions on properties of scalar
mesons, in particular on the production of them in exclusive heavy
flavor hadron decays \cite{wangwei,cheng} have large uncertainties
due to the complicated non-perturbative effects.

It is no doubt that the hadronic light-cone distribution amplitudes
  are the important ingredients when applying factorization
theorem to analyze these exclusive processes. The distribution
amplitudes which are governed by the renormalization group equation
can be obtained by integrating out the transverse momenta of quarks
in hadron for hadronic wave functions. Unfortunately, only twist-2
light cone distribution amplitudes of scalar mesons have been
calculated in Ref.~\cite{cheng} in the framework of QCD sum rules
\cite{shifman}. So the unknown twist-3 distribution amplitudes will
bring obvious uncertainties to the final results. In this work,  we
investigate twist-3 distribution amplitudes of scalar mesons in
order to improve the accuracy of theoretical predictions  of the
scalar mesons.

The calculation of moments for distribution amplitudes making use of
QCD sum rules was presented in much detail in the pioneer work of
\cite{wf method}. Once the moments are known, we can construct
various models to obtain the distribution amplitudes for hadrons.
Following the same method, we will calculate the first two non-zero
moments of twist-3 distribution amplitudes for $a_0(\bar{u}d)$,
$K^*_0(\bar{d}s)$ and $f_0(\bar{s}s)$ respectively based on
renormalization group improved QCD sum rules. Besides, we will
expand the twist-3 distribution amplitudes of scalar mesons
according to Gegenbauer polynomials as usual and use the moments
obtained to determine the first two  Gegenbauer coefficients. As for
$a_0$ and $f_0$ meson, the odd moments for both of the two twist-3
distribution amplitudes (see definition in Eq.~(\ref{wf})) are zero
due to conservation of charge parity and isospin symmetry. However,
the odd moments for $K^*_0$ meson do not vanish when including SU(3)
symmetry breaking effects.

The structure of this paper is  as below: After this introduction,
we derive the general sum rules of moments for twist-3 distribution
amplitudes of scalar mesons in section \ref{form}. Then we will give
the inputs used in our work and present the numerical results of the
first two moments for the above three scalar mesons in section
\ref{resu}. The last section is devoted to our conclusions.

\section{Formulation}
\label{form}

In the valence quark model, there are two  twist-3 light-cone
distribution amplitudes for scalar mesons which are defined as
\cite{cheng}
\begin{eqnarray}
\langle S(p)|\bar{q}_2(y)q_1(x)|0\rangle &=& m_S \bar{f}_{S}\int_0^1
du e^{i(u
p\cdot y+ \bar{u}p \cdot x)} \phi_{S}^{s}(u, \mu), \nonumber \\
\langle S(p)|\bar{q}_2(y) \sigma_{\mu \nu} q_1(x)|0\rangle &=&
-m_{S}\bar{f}_{S}(p_{\mu}z_{\nu}-p_{\nu}z_{\mu})\int_0^1 du e^{i(u
p\cdot y+ \bar{u}p \cdot x)} {\phi_{S}^{\sigma}(u, \mu) \over 6},
\label{wf}
\end{eqnarray}
with $z=y-x$, $\bar{u}=1-u$ and $u$ being the momentum fraction
carried by the $q_2$ quark in the scalar meson. Here the scalar
density meson decay constant $\bar{f}_{S}$ is defined by $\langle
S(p)|\bar{q}_2 q_1|0\rangle=m_S \bar{f}_S$. This scalar decay
constant $\bar{f}_S$ can be connected with the vector current decay
constant $f_S$ which is defined as $\langle
S(p)|\bar{q}_2\gamma_{\mu}q_1|0\rangle=f_S p_{\mu}$:
\begin{eqnarray}
\mu_S f_S= \bar{f}_S, \qquad \mu_S={m_S \over m_2(\mu)-m_1(\mu)} .
\end{eqnarray}
with $m_1$, $m_2$ and $m_S$ being the mass of $q_1$, $q_2$ and
scalar meson respectively. The normalization of these two twist-3
light cone distribution amplitudes are $\int_0^1 du
\phi_{S}^{s}(u)=\int_0^1 du \phi_{S}^{\sigma}(u)=1$. In general,
they have the following form
\begin{eqnarray}
\phi_{S}^{s}(u,\mu)&=&1+\sum_{m=1}^{\infty} a_{m}(\mu)
C_m^{1/2}(2 u-1), \label{an} \\
\phi_{S}^{\sigma}(u,\mu)&=&6 u (1-u) \left[1+\sum_{m=1}^{\infty}
b_{m}(\mu) C_m^{3/2}(2 u-1)\right],\label{bn}
\end{eqnarray}
with the Gegenbauer polynomials $C_1^{1/2}(t)= t$, $ C_2^{1/2}(t)={1
\over 2}(3 t^2-1)$, $C_4^{1/2}(t)={1 \over 8}(35 t^4 -30 t^2 +3), \,
\, C_1^{3/2}(t)=3 t$, $C_2^{3/2}(t)={3 \over 2}(5 t^2-1)$,
$C_4^{3/2}(t)={15 \over 8}(21 t^4 -14 t^2 +1)$, etc.

Now we are ready to calculate the moments of the two twist-3
distribution amplitudes defined in Eq. (\ref{wf}) making use of
background field method in QCD \cite{background method 1,background
method 2, background method 3}. From Eq. (\ref{wf}), one can easily
find
\begin{eqnarray}
\langle 0|\bar{q}_1(0)(i z \cdot \buildrel\leftrightarrow\over
D)^n q_2(0)|S(p)\rangle = m_S \bar{f}_{S}(p \cdot z)^n  \langle \xi_s^n \rangle , \nonumber \\
\langle 0|\bar{q}_1(0)(i z \cdot \buildrel\leftrightarrow\over
D)^{n+1} \sigma_{\mu \nu} q_2(0)|S(p)\rangle = -i {n+1 \over 3}m_S
\bar{f}_{S} (p_{\mu}z_{\nu}-p_{\nu}z_{\mu})(p \cdot z)^n \langle
\xi_{\sigma}^n \rangle,
\end{eqnarray}
with
\begin{eqnarray}
\langle \xi_s^n\rangle=\int_0^1 du (2u-1)^n \phi_S^s(u,\mu), \qquad
\langle \xi^n_{\sigma}\rangle=\int_0^1 du (2u-1)^n
\phi_S^{\sigma}(u,\mu).\label{5}
\end{eqnarray}

In order to calculate the above scalar  moments $\langle
\xi_s^n\rangle$ and   tensor moments $\langle
\xi^n_{\sigma}\rangle$, we consider the following two different
correlation functions, respectively
\begin{eqnarray}
i \int d^4 x e^{i q \cdot x }\langle 0 |T\{\bar{q}_1(x)(i z \cdot
\buildrel\leftrightarrow\over D)^n q_2(x), \bar{q}_2(0) q_1(0)\}|0
\rangle=-(z \cdot q)^n I_s^{(n,0)}(q^2), \label{corre1} \\
i \int d^4 x e^{i q \cdot x }\langle 0 |T\{\bar{q}_1(x)\sigma_{\mu
\nu}(i z \cdot \buildrel\leftrightarrow\over D)^{n+1} q_2(x),
\bar{q}_2(0) q_1(0)\}|0\rangle = i (q_{\mu}z_{\nu}-q_{\nu}z_{\mu})(z
\cdot q)^n I_{\sigma}^{(n,0)} (q^2). \label{correlation function}
\end{eqnarray}

In the deep Euclidean region ($-q^2 \gg 0$), the correlation
functions (\ref{corre1},\ref{correlation function}) can be computed
using operator product expansion at quark level. The results with
power corrections to operators up to dimension-six and lowest order
of $\alpha_s$ corrections are displayed as:
\begin{eqnarray}
{I_s^{(2n,0)}(q^2)}_{QCD}&=&-{3 \over 8 \pi^2} {1 \over 2n+1}
{\rm{ln}}{-q^2 \over \mu^2}(2m_1 m_2 - q^2)+ {\alpha_s \over 8 \pi}
{1 \over q^2} \langle0|G^2|0\rangle \nonumber \\
&&+ {1 \over q^2}\left[({2n+1 \over 2}m_1 +m_2) \langle \bar{q}_1
q_1\rangle + ({2n+1 \over 2}m_2 +m_1)
 \langle \bar{q}_2 q_2\rangle\right]\nonumber
  \\
&& +{1 \over 2} g_s {1 \over q^4}\Bigg\{\left[m_2+n({8 n+11 \over
6}m_1 +2 m_2)\right]\langle \bar{q}_1 \sigma G q_1\rangle \nonumber
\\&&+
\left[m_1+n({8 n+11 \over 6}m_2 +2 m_1)\right]\langle \bar{q}_2
\sigma G q_2\rangle \Bigg\}\nonumber
\\
&& -{4 \pi \alpha_s \over 81}(8 n^2-16 n-21) {\langle \bar{q}_1
q_1\rangle^2 + \langle \bar{q}_2 q_2\rangle^2 \over q^4} + {48 \pi
\alpha_s \over 9}{\langle \bar{q}_1 q_1\rangle \langle \bar{q}_2
q_2\rangle \over q^4} , \label{QCD1}
\end{eqnarray}
for the scalar density even moments,
\begin{eqnarray}
{I_s^{(1,0)}(q^2)}_{QCD}&=& -{3 \over 8 \pi^2}
(m_1^2-m_2^2){\rm{ln}}{-q^2 \over \mu^2}+{\alpha_s \over 4
\pi}(m_1^2-m_2^2)\left[{\rm{ln}}{-q^2 \over \mu^2}+\gamma_E\right]
{1 \over q^4} \langle0|G^2|0\rangle \nonumber
\\
&&+(m_1+m_2){\langle \bar{q}_1 q_1\rangle - \langle \bar{q}_2
q_2\rangle \over q^2} + {10 \pi \alpha_s \over 9}{\langle {\bar{q}_1
q_1\rangle^2 - \langle \bar{q}_2
q_2\rangle^2} \over q^4} \nonumber \\
&&+{g_s \over 2}{1 \over q^4} \left\{({5 \over 4} m_1+2m_2)\langle
\bar{q}_1 \sigma G q_1 \rangle -({5 \over 4} m_2+2m_1)\langle
\bar{q}_2 \sigma G q_2 \rangle \right\} , \label{QCD2}
\end{eqnarray}
for the  first moment of scalar density, and
\begin{eqnarray}
{I_{\sigma}^{(2n,0)}(q^2)}_{QCD}&=&{3 \over 16 \pi^2}{q^2 \over {2
n+3}}\left[1+ 2 \ln{-q^2 \over \mu^2}\right]+ {2n+1 \over 2}{1 \over
q^2}(m_1 \langle \bar{q}_1 q_1\rangle +m_2 \langle \bar{q}_2
q_2\rangle )\nonumber
\\
&& +{16 n^2+14 n+5 \over 24}{g_s \over q^4}[m_1 \langle \bar{q}_1
\sigma G q_1\rangle +m_2 \langle \bar{q}_2 \sigma G q_2\rangle]
\nonumber
\\
&&-{32 n^2 +18 n-35 \over 81}{\pi \alpha_s \over q^4}({\langle
\bar{q}_1
q_1\rangle^2 + \langle \bar{q}_2 q_2\rangle^2}) \nonumber \\
&&+{\alpha_s \over 24 \pi}{1 \over q^2}\langle0|G^2|0\rangle-{8 \pi
\alpha_s \over 9}{1 \over q^4}\langle \bar{q}_1 q_1\rangle \langle
\bar{q}_2 q_2\rangle \delta_{n,0} , \label{QCD3}
\end{eqnarray}
for the tensor even moments, and
\begin{eqnarray}
{I_{\sigma}^{(1,0)}(q^2)}_{QCD}&=&{\alpha_s \over 6
\pi}(m_1^2-m_2^2)\left[2 \ln{-q^2 \over \mu^2} +2 \gamma_E
-3\right]{\langle0|G^2|0\rangle \over q^4} \nonumber \\
&&-{1 \over 4 \pi^2}(m_1^2 - m_2^2)\left[1+ \ln{-q^2 \over
\mu^2}\right]+{3 \over 4}g_s {m_1 \langle \bar{q}_1 \sigma G
q_1\rangle -m_2 \langle \bar{q}_2 \sigma G q_2\rangle \over q^4}\nonumber
 \\
&&+{m_1 \langle \bar{q}_1 q_1\rangle -m_2 \langle \bar{q}_2
q_2\rangle \over q^2}+{2 \pi \alpha_s \over 9}{\langle \bar{q}_1
q_1\rangle^2 - \langle \bar{q}_2 q_2\rangle^2 \over q^4},
\label{QCD4}
\end{eqnarray}
for the  first moment of tensor sum rule. Since we are concerned
with only the first two Gegenbauer coefficients, we do not display
the explicit forms of sum rules for other odd moments for
simplification.
 When $n$ is equal to 0, the Eq.~(\ref{QCD1}) is in accord
with the results shown in Ref. \cite{reinders}. On the other hand,
the correlation functions (\ref{corre1},\ref{correlation function})
can also be calculated at hadron level by inserting a complete set
of quantum states $\Sigma |n\rangle \langle n|$, which are written
as
\begin{eqnarray}
&&{\mbox{Im} I_s^{(2n,0)}(q^2)}_{had}=-\pi \delta(q^2-m_S^2) m_S^2
\bar{f}_S^2 \langle \xi_s^{2n} \rangle +\pi {3 \over 8 \pi^2}{1
\over
2n+1}(2 m_1 m_2-q^2)\theta(q^2-s_S), \,\,\, \label{had 1}\\
&&{\mbox{Im} I_s^{(1,0)}(q^2)}_{had}=-\pi \delta(q^2-m_S^2) m_S^2
\bar{f}_S^2 \langle \xi_s^{1} \rangle + \pi {3 \over 8 \pi^2}
(m_1^2-m_2^2)\theta(q^2-s_S), \label{had2}\\
&&{\mbox{Im}I_{\sigma}^{(2n,0)}(q^2)}_{had}=-\pi \delta(q^2-m_S^2)
{2n+1 \over 3}m_S^2 \bar{f}_S^2 \langle \xi_{\sigma}^{2n} \rangle -
\pi {3 \over 8 \pi^2}{1 \over 2n+3} q^2 \theta(q^2-s^{\sigma}_{S}),
\label{had3}\\
&&{\mbox{Im} I_{\sigma}^{(1,0)}(q^2)}_{had}=-\pi \delta(q^2-m_S^2)
{2 \over 3}m_S^2 \bar{f}_S^2 \langle \xi_{\sigma}^1 \rangle +\pi {1
\over 4 \pi^2}(m_1^2-m_2^2) \theta(q^2-s^{\sigma}_{S}).\label{had 4}
\end{eqnarray}
Here the quark-hadron duality has been used to obtain the above
equations. Then we can match these two different representations of
correlation functions (\ref{corre1},\ref{correlation function})
calculated in quark and hadron level by the dispersion relation
\begin{eqnarray}
{1 \over \pi} \int ds {\mbox{Im} I(s)_{had} \over
s-q^2}=I(q^2)_{QCD}.
\end{eqnarray}

In order to suppress the contributions from the excited resonances
and continuum states, we apply the Borel transformation to both
sides of the above equation. On the other hand, this transformation
can also remove the arbitrary polynomials in $q^2$. Then we obtain
\begin{eqnarray}
{1 \over \pi }{1 \over M^2}\int ds \, e^{-s/M^2} \mbox{Im}
I(s)_{had}=\mathcal{B}_{M^2} I(q^2)_{QCD}, \label{dispersion
relation}
\end{eqnarray}
where $M$ is the Borel parameter, and $\mathcal{B}_{M^2}$ is the
operator of Borel transformation which is defined as
\cite{shifman,practioner 1}
\begin{eqnarray}
\mathcal{B}_{M^2}=\lim_{\stackrel{-q^2,n \to \infty}{-q^2/n=M^2}}
\frac{(-q^2)^{(n+1)}}{n!}\left( \frac{d}{dq^2}\right)^n.
\end{eqnarray}

Finally, substituting Eqs.~(\ref{QCD1}-\ref{QCD4}) and (\ref{had
1}-\ref{had 4}) into the Eq.~(\ref{dispersion relation}), we have
the scalar density even moments
\begin{eqnarray}
-m_S^2 \bar{f}_S^2 e^{-m_S^2/M^2}\langle \xi^{2n}_s\rangle&=&{3
\over 8 \pi^2}{1 \over 2n+1}\int^{s_{S}}_0 (2m_1 m_2-s)e^{-s/M^2} ds
-{\alpha_s \over 8 \pi}\langle 0 |G^2|0 \rangle
\nonumber \\
&& -\left[\left({2n+1 \over 2}m_1 +m_2\right)\langle \bar{q}_1
q_1\rangle +\left({2n+1 \over 2}m_2 +m_1\right)\langle \bar{q}_2
q_2\rangle\right]
\nonumber \\
&&+{g_s \over 2M^2} \bigg \{\left[m_2+n\left({8n+11 \over 6} m_1 +2
m_2\right)\right]\langle \bar{q}_1 \sigma G q_1\rangle +\nonumber
\\
&&\left[m_1+n\left({8n+11 \over 6} m_2 +2 m_1\right)\right]\langle
\bar{q}_2 \sigma G q_2\rangle \bigg \}+\frac{48 \pi \alpha_s}{   9
M^2}\langle \bar{q}_1 q_1\rangle \langle \bar{q}_2 q_2\rangle
\nonumber
\\
&&-{4 \pi \alpha_s \over 81}(8n^2-16n-21){1 \over M^2}\left(\langle
\bar{q}_1 q_1\rangle^2+\langle \bar{q}_2 q_2\rangle^2\right),
\label{SR1}
\end{eqnarray}
the  first moment of scalar density
\begin{eqnarray}
-m_S^2\bar{f}_S^2 e^{-m_S^2/M^2}\langle \xi^{1}_s\rangle&=& {3 \over
8 \pi^2}(m_1^2-m_2^2) \int_0^{s_{S}} d s
e^{-s/M^2}-(m_1+m_2)(\langle \bar{q}_1 q_1\rangle-\langle \bar{q}_2
q_2\rangle)
\nonumber\\
&&+ (m_1^2-m_2^2){1-\ln{\mu^2 \over M^2} \over 4M^2}\langle
{\alpha_s \over \pi} G^2\rangle +{10 \pi \alpha_s \over 9M^2}
\left(\langle \bar{q}_1 q_1\rangle^2-\langle \bar{q}_2
q_2\rangle^2\right)
\nonumber  \\
&&+{g_s \over 2M^2} \left[\left({5 \over 4} m_1 +2 m_2\right)\langle
\bar{q}_1 \sigma G q_1\rangle-\left({5 \over 4} m_2 +2
m_1\right)\langle \bar{q}_2 \sigma G q_2\rangle\right],
\,\,\,\label{SR2}
\end{eqnarray}
tensor even moments
\begin{eqnarray}
-{2n+1 \over 3}m_S^2 \bar{f}_S^2 e^{-m_S^2/M^2}\langle
\xi_{\sigma}^{2n}\rangle &=&-{3 \over 8 \pi^2}{1 \over
2n+3}\int^{s^{\sigma}_S}_0 s e^{-s/M^2}ds -{8 \pi \alpha_s \over
9}{\langle \bar{q}_1 q_1\rangle \langle \bar{q}_2 q_2\rangle \over
M^2}\delta_{n,0}
\nonumber \\
&&-{2n+1 \over 2}\left(m_1 \langle \bar{q}_1 q_1\rangle+ m_2 \langle
\bar{q}_2 q_2\rangle\right )- {\alpha_s \over 24
\pi}\langle0|G^2|0\rangle
\nonumber \\
&& +{16 n^2+14n+5 \over 24}g_s {m_1 \langle \bar{q}_1 \sigma G
q_1\rangle + m_2 \langle \bar{q}_2 \sigma G q_2\rangle \over M^2}
\nonumber \\
&&-{(32n^2 +18n-35)\pi \alpha_s \over 81}{\langle \bar{q}_1
q_1\rangle^2+\langle \bar{q}_2 q_2\rangle^2 \over M^2}, \label{SR3}
\end{eqnarray}
and the  first moment of tensor current
\begin{eqnarray} -{2 \over 3}m_S^2 \bar{f}_S^2
e^{-m_S^2/M^2}\langle \xi_{\sigma}^{1}\rangle &=& {1 \over 4
\pi^2}(m_1^2-m_2^2)\int _0^{s^{\sigma}_S} e^{-s/M^2}ds -(m_1 \langle
\bar{q}_1 q_1\rangle -
m_2 \langle \bar{q}_2 q_2\rangle)\nonumber \\
&& -{\alpha_s \over 6 \pi}(m_1^2-m_2^2){1+2 \ln {\mu^2 \over M^2}
\over M^2} \langle0|G^2|0\rangle
\nonumber \\
&&+{3 \over 4}g_s {m_1 \langle \bar{q}_1 \sigma G q_1\rangle - m_2
\langle \bar{q}_2 \sigma G q_2\rangle \over M^2} +{2 \pi \alpha_s
\over 9}{\langle \bar{q}_1 q_1\rangle^2-\langle \bar{q}_2
q_2\rangle^2 \over M^2}. \label{SR4}
\end{eqnarray}
Here the vacuum saturation approximation \cite{practioner
1,practioner 2} has been used to describe the four quark condensate,
i.e.,
\begin{eqnarray}
&&\langle 0| \bar{q}^A_{\alpha a}(x) \bar{q}^B_{\beta
b}(y)q^C_{\gamma c}q^D_{\delta d}|0\rangle \nonumber
 \\
&&={1 \over 144}\left[\delta_{AD}\delta_{BC} \delta_{\alpha \delta}
\delta_{\beta \gamma} \delta_{ad} \delta_{bc}-\delta_{AC}\delta_{BD}
\delta_{\alpha \gamma} \delta_{\beta \delta} \delta_{ac}
\delta_{bd}\right]\langle \bar{q}^A q^A \rangle \langle \bar{q}^B
q^B \rangle.
\end{eqnarray}
Here $\alpha,\beta,\gamma,\delta$ are the spinor indices, $a,b,c,d$
are the color indices, and $A,B,C,D$ denote the flavor of quarks.
Besides, the flavor indices in the right hand side of the above
equation do not mean the sum of all flavors.

It is noted that all the parameters in the above sum rules are fixed
at scale of Borel mass $M$. The renormalization group   equations of
decay constant, quark mass and condensate are given as \cite{yang}
\begin{eqnarray}
&&\bar{f}_S(M)=\bar{f}_S(\mu) \bigg({\alpha_s (\mu) \over \alpha_s
(M)}\bigg)^{4/b}, \qquad  m_{q,M}=m_{q,{\mu}} \bigg({\alpha_s (\mu)
\over \alpha_s (M)}\bigg)^{-4/b}, \nonumber \\
&&\langle \bar{q} q\rangle_{M}= \langle \bar{q} q\rangle_{\mu}
\bigg({\alpha_s (\mu) \over \alpha_s (M)}\bigg)^{4/b}, \qquad
\langle g_s \bar{q} \sigma   G q\rangle_{M}= \langle g_s \bar{q}
\sigma G q\rangle_{\mu} \bigg({\alpha_s (\mu) \over \alpha_s
(M)}\bigg)^{-2/3b},  \nonumber\\
&&\langle \alpha_s G^2 \rangle_{M} =\langle \alpha_s G^2
\rangle_{\mu},  \label{RG}
\end{eqnarray}
 with $b= (33 -2 n_f)/3$, $n_f$ is the number of active quark flavors.
Making use of the orthogonality of Gegenbauer polynomials
\begin{eqnarray}
&&\int^1_0  dx C_n^{1/2}(2x-1)C_m^{1/2}(2x-1)={1\over
2 n+1}\delta_{m n}, \nonumber \\
&&\int^1_0  dx x(1-x)C_n^{3/2}(2x-1)C_m^{3/2}(2x-1)={(n+2)(n+1)\over
4 (2n+3)}\delta_{m n},
\end{eqnarray}
the Gegenbauer moments $a_l$, $b_l$ can be related to moments,
$\langle \xi^k_s \rangle$, $\langle \xi^k_{\sigma} \rangle$
for example:
\begin{eqnarray}
&&a_1=3 \langle \xi_1\rangle, \qquad a_2={5\over 2}( 3 \langle \xi_2\rangle-1), \qquad a_4={9\over 8}( 35 \langle \xi_4 \rangle-30 \langle \xi_2 \rangle+3), \nonumber\\
&&b_1={5 \over 3} \langle \xi_1 \rangle, \qquad b_2={7 \over 12}( 5
\langle \xi_2\rangle -1), \qquad b_4={11\over 24} (21 \langle
\xi_4\rangle -14 \langle \xi_2 \rangle+1).\label{gege}
\end{eqnarray}
The renormalization group  equations of Gegenbauer moments are given
as
\begin{eqnarray}
\langle a_n (\mu)\rangle=\langle a_n
(\mu_0)\rangle\bigg({\alpha_s(\mu_0) \over
\alpha_s(\mu)}\bigg)^{-\gamma^S_n /b}, \qquad \langle b_n
(\mu)\rangle=\langle b_n(\mu_0)\rangle\bigg({\alpha_s(\mu_0) \over
\alpha_s(\mu)}\bigg)^{-\gamma^T_n /b}, \label{momentsRG1}
\end{eqnarray}
where the one-loop anomalous dimensions are\cite{shifman RG}
\begin{eqnarray}
\gamma^S_n=C_F \bigg(1- {8 \over (n+1)(n+2)}+4 \sum_{j=2}^{n+1}
\frac{1}{j} \bigg), \qquad  \gamma^T_n=C_F \bigg(1+4
\sum_{j=2}^{n+2} \frac{1}{j}\bigg), \label{momentsRG2}
\end{eqnarray}
with $C_F=4/3$.

\section{Numerical results and discussions}
\label{resu}

The input parameters used in this paper are taken as
\cite{ioffe,cheng,wuxh,practioner 1}

\begin{equation}
\begin{array}{ll}
 \langle \bar{s} s\rangle = (0.8 \pm 0.1) \langle \bar{u} u\rangle, & \langle \bar{u} u\rangle \cong
\langle \bar{d} d\rangle \cong -(1.65 \pm 0.15) \times
10^{-2}{\rm{GeV}}^3,
 \\
\langle {\alpha_s \over \pi} G_{\mu \nu}^a G^{a \mu
\nu}\rangle=(0.005 \pm 0.004) {\rm{GeV}}^4, & \langle g_s
\bar{u}\sigma G u\rangle \cong \langle g_s \bar{d}\sigma G d\rangle
=m_0^2 \langle \bar{u}u\rangle,
 \\
  m_u(1 {\rm{GeV}})=2.8 {\rm{MeV}},
&  \langle g_s \bar{s}\sigma G s\rangle =(0.8 \pm 0.1) \langle g_s
\bar{u}\sigma
G u\rangle,\\
m_d(1 {\rm{GeV}})=6.8 {\rm{MeV}}, & \alpha_{s}(1 {\rm{GeV}})=0.517,\\
m_s(1 {\rm{GeV}})=142 {\rm{MeV}}, & m_0^2=(0.8 \pm 0.2) {\rm{GeV}}
^2.
\end{array}
\end{equation}
Here all the values for vacuum condensates are adopted at the scale
$\mu= 1 {\rm{GeV}}$. Next we are ready to analyze the sum rules for
the scalar meson nonet one by one.

\subsection{Mass, decay constant and moments for $a_0$ meson}

\subsubsection{
 Determination of mass, decay constant and scalar moments $\langle
\xi^{2(4)}_{s,a_0} \rangle$ of $a_0$ from sum rules in (\ref{SR1}) }

\begin{figure}[tb]
\begin{center}
\begin{tabular}{ccc}
\includegraphics[scale=1.0]{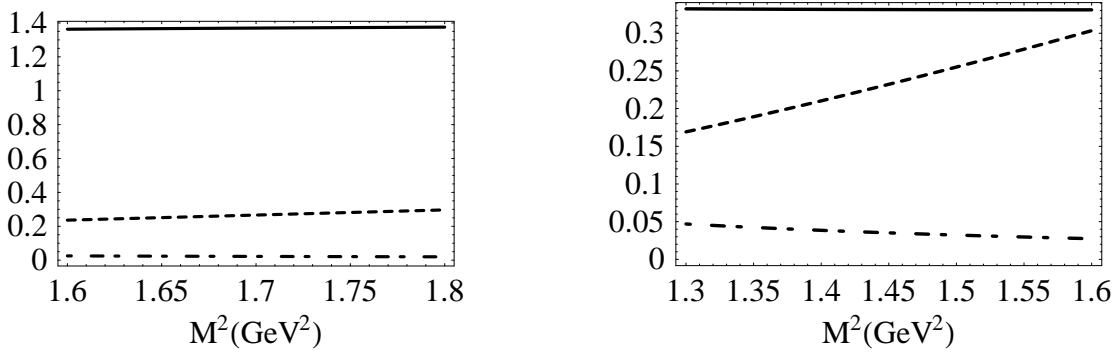}
\end{tabular}
\caption{Mass (left solid line) and decay constant (right solid
line) of $a_0$ from scalar sum rule in Eq.~(\ref{SR1}) with $s_S=4.5
\,\,{\rm{GeV^2}}$ as a function of Borel parameter $M^2$. The dashed
line denotes the contribution from continuum states in the total sum
rules and the dot-dashed line is the ratio of dimension-six
condensate contribution in the total sum rules.}\label{massa01}
\end{center}
\end{figure}

Here $a_0$ is the scalar meson with quark contents $\bar{q}_1 q_2=
\bar{d} u$. A common way to obtain the sum rules of meson mass from
Eq.~(\ref{SR1}) is taking the logarithm of both sides of this
equation, and then applying the differential operator $M^4
\partial/ \partial M^2$ to them. However, firstly we need to fix
the value of threshold parameter and Borel parameter in order to
obtain the value of the mass. As far as the threshold parameter
$s_S$ is concerned, its value should be adopted so that the Borel
window is stable enough which indicates that the mass is independent
of the choice for $M^2$ in some region. For the choice of the Borel
parameter, one requires that the contribution from continuum states
is less than 30\% and the contribution from dimension-six
condensates is less than 10\%. In view of the above requirements, we
choose the threshold parameter $s_S=(4.5 \pm 0.3)\,\,{\rm{GeV}^2}$,
such that the stable Borel window is in the range $M^2 \in [1.60,
1.80] \,\,{\rm{GeV}^2}$ which is shown in Fig.~\ref{massa01}. From
this figure, we can observe that the mass for $\bar{q}_1 q_2=
\bar{d} u$ scalar ground state is $m_{a_0}=(1320 \sim
1410)\,\,{\rm{MeV}}$. This is very close to the physical state $a_0
(1450)$ \cite{pdg}. It should be pointed out that the possibility of
the existence of light scalar resonance near 1.4 GeV was firstly
predicted by Ref. \cite{kataev} as the first radial excitations of
$a_0(980)$ according to the so called "linear dual models" on the
assumptions of $\bar{q}q$ structure of $a_0(980)$. The decay
constant of $a_0$ can be easily read from the sum rules in
Eq.~(\ref{SR1}) as soon as the mass is known. The decay constant
within the Borel window is also plotted in Fig.~\ref{massa01} as the
second diagram. It is easy to find that the decay constant is quite
stable within the Borel window $M^2 \in [1.30, 1.60]
\,\,{\rm{GeV}^2}$ when the contribution from continuum states and
the dimension-six condensate is less than 30\% and 10\%,
respectively. Therefore, we obtain the decay constant as $
\bar{f}_{a_0}{\rm{(1 GeV)}}=(322 \sim 341)\,\,{\rm{MeV}}$. In the
following subsections, all the values of decay constants and moments
are calculated at scale of 1 GeV unless explicitly pointed out.

\begin{figure}[tb]
\begin{center}
\begin{tabular}{ccc}
\includegraphics[scale=1.0]{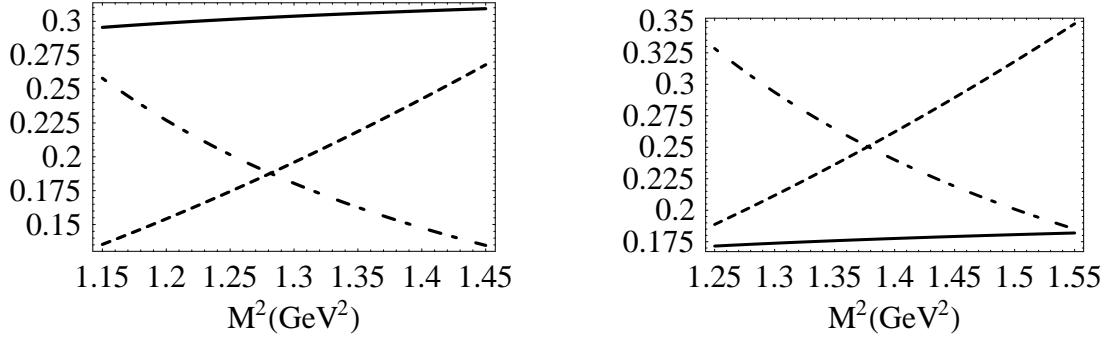}
\end{tabular}
\caption{$\langle \xi^{2}_{s,a_0} \rangle$ (left solid line) and
$\langle \xi^{4}_{s,a_0} \rangle$ (right solid line) from scalar sum
rules in Eq.~(\ref{SR1}) with $s_S=4.5\,\,{\rm{GeV^2}}$ as a
function of Borel parameter $M^2$. The dashed and the dot-dashed
lines are the ratio of contribution from continuum states and
dimension-six condensates, respectively.}\label{momenta01}
\end{center}
\end{figure}

From the definition of twist-3 distribution amplitudes for $a_0$, it
can be found that only even Gegenbauer moments are non-zero due to
conservation of charge parity and isospin symmetry as mentioned in
the introduction. Next we are going to consider the second and
fourth moments of $\phi_{a_0}^s$ from scalar density sum rules for
$a_0$ meson. Just as the determination of mass and decay constant,
one should find a stable window for the sum rule of each moment. The
contributions of continuum states and dimension-six condensates  are
plotted in Fig.~\ref{momenta01},  where the moments $\langle
\xi^{2(4)}_{s,a_0} \rangle$ within the Borel window $M^2 \in [1.15,
1.45]\,\,$GeV$^2$ ($[1.25, 1.55]\,\,$GeV$^2$) are also included. For
the second (fourth) moments, the contributions from both continuum
states and the dimension-six condensates are less than 30\% (35\%).
Then we have $\langle \xi^{2}_{s,a_0} \rangle=0.29 \sim 0.31$ and
$\langle \xi^{4}_{s,a_0} \rangle=0.16 \sim 0.19$.

\subsubsection{
  Determination of mass, decay constant and tensor moments
$\langle \xi^{2(4)}_{\sigma,a_0} \rangle$ of $a_0$ from sum rules in
(\ref{SR3})}

\begin{figure}[tb]
\begin{center}
\begin{tabular}{ccc}
\includegraphics[scale=1.0]{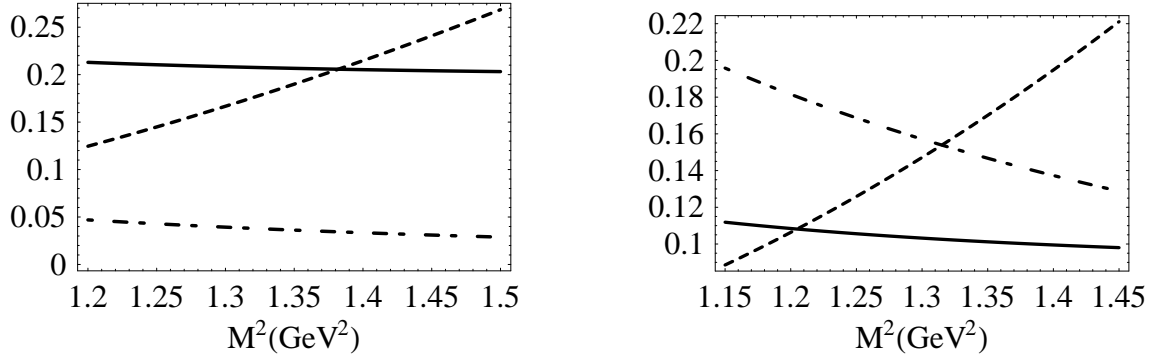}
\end{tabular}
 \caption{$\langle \xi^{2}_{\sigma,a_0}\rangle$ (left solid line) and
$\langle \xi^{4}_{\sigma,a_0} \rangle$ (right solid line) from
tensor sum rules in Eq.~(\ref{SR3}) with $s_S^\sigma=4.5
\,\,{\rm{GeV^2}}$ as a function of Borel parameter $M^2$. The dashed
and the dot-dashed line represent the ratio of contribution from the
continuum states and dimension-six condensates.} \label{momenta02}
\end{center}
\end{figure}

In the above, we have got the mass and decay constant for $a_0$
meson from scalar density sum rules in Eq.~(\ref{SR1}). Similarly,
we can also extract them from tensor sum rules in Eq.~(\ref{SR3}).
Moreover, the values of mass and decay constant may  not be exactly
the same between these two sum rules due to different correlation
functions adopted for them. Following the similar procedure, one can
get the mass and decay constant: $m_{a_0}=(1270 \sim 1390)
\,\,{\rm{MeV}}, \bar{f}_{a_0}=(325 \sim 350)\,\, \rm{MeV}$, which
are very close to the range we got from the sum rules of
Eq.(\ref{SR1}) in previous subsection. The mass (decay constant) is
obtained under the condition that the contributions from both
continuum states and the dimension-six condensates should be less
than 30\% (25 \%) respectively in total sum rules. The threshold
parameter $s^{\sigma}_{S}$ is still adopted as $(4.5 \pm 0.3)
\,\,{\rm{GeV}^2}$, while the Borel windows are $M^2 \in [1.60, 1.80]
\,\,{\rm{GeV}^2}$ and $[1.20, 1.50] \,\,{\rm{GeV}^2}$, respectively.
Making use of the mass and decay constant, we can determine the
second and fourth moments $\langle \xi^{2}_{\sigma,a_0}\rangle$,
$\langle \xi^{4}_{\sigma,a_0}\rangle$ for the tensor twist-3
distribution amplitude of $a_0$ meson within the Borel window $M^2
\in [1.20, 1.50] \,\,{\rm{GeV}^2}$ and $[1.15, 1.45]
\,\,{\rm{GeV}^2}$ as shown in Fig.\ref{momenta02}. Here the
contributions from continuum states and the dimension-six condensate
are no more than 30\%, which indicate that the sum rules for these
two moments are reliable. Hence, the results for $\langle
\xi^{2}_{\sigma,a_0}\rangle$ and $\langle
\xi^{4}_{\sigma,a_0}\rangle$ are $ 0.20 \sim 0.22$ and $ 0.093 \sim
0.12$, respectively, within the given Borel window and threshold
parameter.

\subsection{Mass, decay constant and moments for $K^{*}_0$ meson}

\subsubsection{ Determination of mass, decay constant and moments
$\langle \xi^{1(2)}_{s,K^{*}_0} \rangle$ of $K^{*}_0$ from scalar
density sum rules     }

As explained before, here the scalar meson $K^{*}_0$ is made up of
$\bar{u}s$ quarks. Different from the $a_0$ meson, both odd and even
moments of distribution amplitudes for $K^{*}_0$ are non-zero. The
mass and decay constant of $K^{*}_0$ can be derived from scalar
density sum rules in Eq.~(\ref{SR1}) following the same method as
done for $a_0$ case. The threshold value is chosen as $s_S=(5.4 \pm
0.3)\,\,{\rm{GeV}^2}$ in the sum rules of Eq.(\ref{SR1}) for
$K^{*}_0$ meson in order to gain the stable Borel window $M^2 \in
[1.90, 2.10] \,\,{\rm{GeV}^2}$ and $ [1.30, 1.70] \,\,{\rm{GeV}^2}$
for mass and decay constant, respectively. Then we can obtain the
value of mass (decay constant) of $K^{*}_0$ as $m_{K^{*}_0}= (1460
\sim 1560)$ {MeV} ($\bar{f}_{K^{*}_0}=(344 \sim 368)\,\, \rm{MeV}$)
with the requirement that the contributions from both continuum
states and dimension-six operator are less than 30\% (25\%).

\begin{figure}[tb]
\begin{center}
\begin{tabular}{ccc}
\includegraphics[scale=1.0]{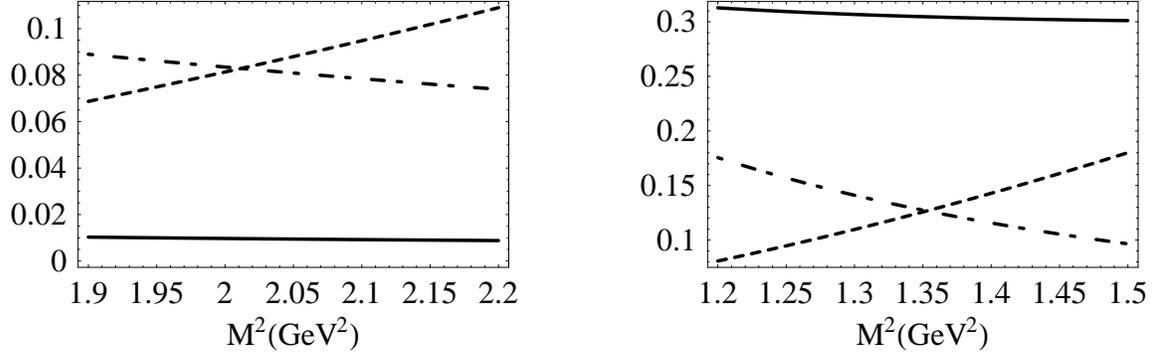}
\end{tabular}
 \caption{$\langle \xi^{1}_{s,K^*_0} \rangle$ (left solid line)  from scalar sum rules
(\ref{SR2}) and $\langle \xi^{2}_{s,K^*_0}\rangle$ (right solid
line) from sum rules (\ref{SR1}) with $s_S=5.4 \,\,{\rm{GeV}^2}$ as
a function of Borel parameter $M^2$. The dashed and the dot-dashed
lines indicate the ratio of continuum states and dimension-six
condensates to the total sum rules, respectively.} \label{momentk01}
\end{center}
\end{figure}

Then we try to calculate the first and second moment for $K^{*}_0$
meson scalar twist-3 distribution amplitude according to sum rules
(\ref{SR2}) and (\ref{SR1}), respectively. For the first moment
$\langle \xi^{1}_{s,K^{*}_0}\rangle$ of scalar density, we require
that the contributions from both the continuum states and
dimension-six condensates should be less than 15\% in order to
obtain stable Borel window. As for the sum rules of the second
moment $\langle \xi^{2}_{s,K^{*}_0} \rangle$, the contributions from
both the continuum states and dimension-six operators are less than
20\%. From the Fig.~\ref{momentk01}, we can read out the results of
the first scalar moment $\langle \xi^{1}_{s,K^{*}_0}\rangle$ as
$(0.61 \sim 1.42) \times 10^{-2}$ within the Borel window $M^2 \in
[1.90, 2.20] \,\,{\rm{GeV}^2}$, and the second moment $\langle
\xi^{2}_{s,K^{*}_0}\rangle$ as  $0.29 \sim 0.33$ within the Borel
window   $[1.20, 1.50] \,\,{\rm{GeV}^2}$. The threshold parameter is
fixed at $s_S=(5.4 \pm 0.3)\,\,{\rm{GeV}^2}$.

\subsubsection{  Determination of mass, decay constant and moments
$\langle \xi^{1(2)}_{\sigma,K^{*}_0} \rangle$ of $K^{*}_0$ from
tensor current sum rules}

\begin{figure}[tb]
\begin{center}
\begin{tabular}{ccc}
\includegraphics[scale=1.0]{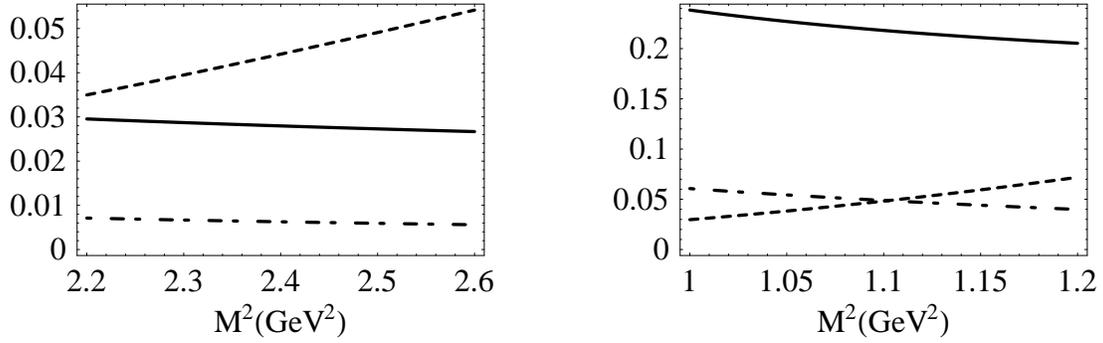}
\end{tabular}
\caption{$\langle \xi^{1}_{\sigma,K^{*}_0}\rangle$ (left solid line)
from tensor sum rules in (\ref{SR4}) and $\langle
\xi^{2}_{\sigma,K^{*}_0}\rangle$ (right solid line) from sum rules
in (\ref{SR3}) with $s_S^{\sigma}=5.4 \,\,{\rm{GeV}^2}$ as a
function of Borel parameter $M^2$. The dashed and the dot-dashed
lines represent the ratio corresponding to the contribution from
continuum states and dimension-six condensates in the total sum
rules, respectively.} \label{momentk02}
\end{center}
\end{figure}

Similarly, we can also derive the results of mass and decay constant
from the tensor operator sum Rules in Eq.~(\ref{SR3}). Here we will
only show our values of them as $m_{K^{*}_0}=(1440 \sim 1550)
\,\,{\rm{MeV}}$ and $\bar{f}_{K^{*}_0}=(349 \sim 375)
\,\,{\rm{MeV}}$ within the Borel window $M^2 \in [2.00, 2.20]
\,\,{\rm{GeV}^2}$ and $ [1.30, 1.60] \,\,{\rm{GeV}^2}$. The
threshold parameter is set the same as before, $s_S^\sigma=(5.4 \pm
0.3)\,\, {\rm{GeV}^2}$. The contributions from both the continuum
states and dimension-six condensates are required to be less than
30\% (20\%) for mass (decay constant) sum rules respectively. The
mass of $K^*_0$ determined here are consistent with the one
determined in the previous subsection from sum rules (\ref{SR1}),
which is quite close to the physical state $K_0^*(1430)$.
 The first and second moments
$\langle \xi^{1}_{\sigma,K^{*}_0}\rangle$, $\langle
\xi^{2}_{\sigma,K^{*}_0}\rangle$ of the tensor twist-3 distribution
amplitude can be computed following the same method, which have been
plotted in Fig.\ref{momentk02}. It is found that the sum rules for
these two moments are quite stable within the Borel window $M^2 \in
[2.20, 2.60] \,\,{\rm{GeV}^2}$ and $[1.00, 1.20] \,\,{\rm{GeV}^2}$,
respectively,  since contributions from both continuum states and
dimension-six condensates are less than 10\%. The results for them
are shown as: $\langle \xi^{1}_{\sigma,K^{*}_0}\rangle= (2.2 \sim
3.3) \times 10^{-2}$, $\langle \xi^{2}_{\sigma,K^{*}_0}\rangle= 0.20
\sim 0.25$.

\subsection{Mass, decay constant and moments for $f_0$ meson}

\subsubsection{
Determination of mass, decay constants and scalar moments $\langle
\xi^{2(4)}_{s,f_0} \rangle$ of $f_0$ from Sum Rules in (\ref{SR1})
}

\begin{figure}[tb]
\begin{center}
\begin{tabular}{ccc}
\includegraphics[scale=1.0]{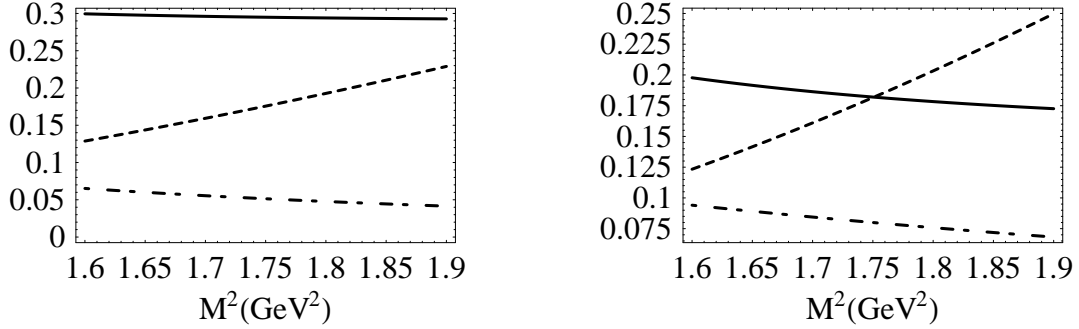}
\end{tabular}
\caption{$\langle \xi^{2}_{s,f_0} \rangle$ (left solid line) and
$\langle \xi^{4}_{s,f_0} \rangle$ (right solid line) from scalar sum
rules in Eq.~(\ref{SR1}) with $s_S=6.5\,\,{\rm{GeV^2}}$ as a
function of Borel parameter $M^2$. The dashed and the dot-dashed
lines reflect the ratio of continuum states and dimension-six
condensates to the total sum rules, respectively.} \label{momentf01}
\end{center}
\end{figure}

Here  $f_0$ refers to the scalar meson which is made up of $\bar{s}
s$ quark. The sum rules for $f_0$ are much the same as for the $a_0$
meson. The odd moments vanish due to conservation of C parity.
Therefore, we will only consider the first two even moments,
$\langle \xi^{2}_{s,f_0} \rangle$ and $\langle \xi^{4}_{s,f_0}
\rangle$ of the scalar twist-3 distribution amplitude for the $f_0$
meson. Since the calculations are similar as that for $a_0$ meson,
the mass and decay constant for $f_0$  are given straightforward as
$m_{f_0}=(1640 \sim 1730) \,\,{\rm{MeV}}$ and $\bar{f}_{f_0}=(369
\sim 391)\,\,{\rm{MeV}}$ within the Borel window $M^2 \in [2.50,
2.70] \,\,{\rm{GeV}^2}$, $ [1.70, 2.00] \,\,{\rm{GeV}^2}$
respectively. The threshold parameter is set as $s_S=(6.5 \pm 0.3)
\,\,{\rm{GeV}}^2$. Here we also require that the contributions from
both the continuum states and dimension-six condensates are less
than 30\% (20\%) for mass (decay constant) sum rules respectively.
As for the second and forth moments $\langle
\xi^{2}_{s,f_0}\rangle$,$\langle \xi^{4}_{s,f_0}\rangle$, the
results within the same Borel window $M^2 \in [1.60, 1.90]
\,\,{\rm{GeV}^2}$ are plotted in Fig.~\ref{momentf01}. The number of
$\langle \xi^{2}_{s,f_0}\rangle$, $\langle \xi^{4}_{s,f_0}\rangle$
are $0.29 \sim 0.31$ and $0.17 \sim 0.20$ respectively within the
given Borel window and threshold parameter. The requirement that the
contributions from the continuum states and dimension-six operator
are less than 25\% (30\%) for the second (forth) scalar moments has
been used.

\subsubsection{ Determination of mass, decay constant and tensor moments $\langle
\xi^{2(4)}_{\sigma,f_0} \rangle$ of $f_0$ from Sum Rules in
(\ref{SR3}) }
\begin{figure}[tb]
\begin{center}
\begin{tabular}{ccc}
\includegraphics[scale=1.0]{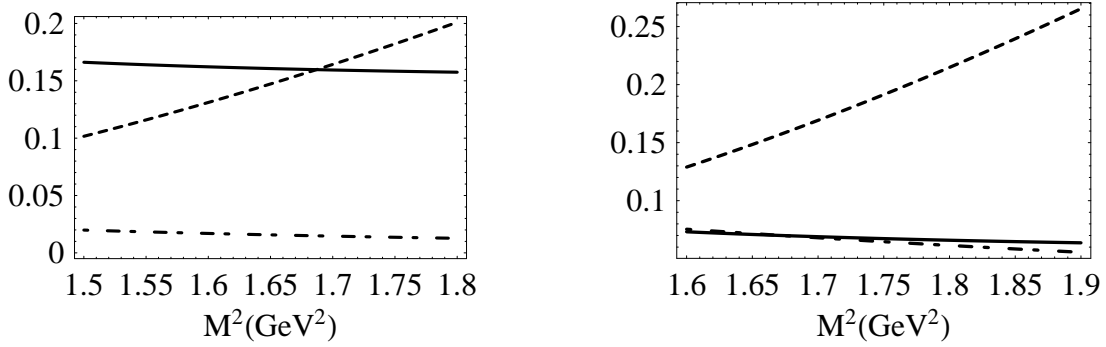}
\end{tabular}
 \caption{$\langle \xi^{2}_{\sigma,f_0}\rangle$ (left solid line) and
$\langle \xi^{4}_{\sigma,f_0} \rangle$ (right solid line) from
tensor sum rules in Eq.~(\ref{SR3}) with $s_S^\sigma=6.5
\,\,{\rm{GeV^2}}$ as a function of Borel parameter $M^2$. The dashed
and the dot-dashed line represent the ratio of contribution from the
continuum states and dimension-six condensates.} \label{momentf02}
\end{center}
\end{figure}
The mass and decay constants of $f_0$ can also be derived from
tensor sum Rules in Eq.~(\ref{SR3}). Adopting the same threshold
parameter as the scalar density sum rules, we obtain the results as
$m_{f_0}= ({1620 \sim 1710) \rm{MeV}}$ and $\bar{f}_{f_0}= (381 \sim
426) {\rm{MeV}}$ within the Borel window $M^2 \in [2.50, 2.70]
\,\,{\rm{GeV}^2}$, $ [1.20, 1.60] \,\,{\rm{GeV}^2}$ respectively.
Here we require that contributions from both continuum states and
dimension-six operators are less than 30\% (10\%) for mass (decay
constant) sum rules. The mass we get here from tensor sum rules and
also that from the scalar density sum rules (\ref{SR1}) in previous
subsection is close to the physical state $f_0 (1710)$. The second
and forth moment $\langle \xi^{2}_{\sigma,f_0} \rangle$,  $\langle
\xi^{4}_{\sigma,f_0} \rangle$ of tensor twist-3 distribution
amplitude are also displayed in Fig.~\ref{momentf02} within the
Borel window $M^2 \in [1.50, 1.80] \,\,{\rm{GeV}^2}$ and $[1.60,
1.90] \,\,{\rm{GeV}^2}$ respectively. The condition that the
contributions from the continuum states and dimension-six operators
are less than 25\% (30\%) is adopted for the second (forth) tensor
moment. The value of $\langle \xi^{2}_{\sigma,f_0}\rangle$ and
$\langle \xi^{4}_{\sigma,f_0}\rangle$ are $0.15 \sim 0.17$ and
$0.057 \sim 0.082$ within the given Borel window and threshold
parameter.

\begin{table}[tb]
\caption{Masses, decay constants and Gegenbauer moments from the
scalar density  sum rules (\ref{SR1},\ref{SR2}) at the scale $\mu=1
{\rm{GeV}}$ and 2.1GeV (shown in the second line of each meson)  }
\begin{center}
\begin{tabular}{|c|c|c|c|c|c|}
\hline\hline
 state & $m${(\rm{MeV})}& $\bar{f}${(\rm{MeV})}& $a_1 (
\times 10^{-2})$ & $a_2$ & $a_4$\\
\hline $a_0$&$1320 \sim 1410$ &$322 \sim 341$&
0 & $-0.33 \sim -0.18 $   & $-0.11 \sim 0.39$  \\
& &$391 \sim 414 $& & $-0.26 \sim -0.14$ & $-0.075 \sim 0.27$\\
\hline$K^*_0$&$1460 \sim 1560$ &$344 \sim 368$& $1.8 \sim 4.2$ & $-0.33 \sim -0.025$ & --- \\
& & $418 \sim 447 $& $1.6 \sim 3.8$& $-0.26 \sim -0.020$ & \\
\hline$f_0$&$1640 \sim 1730$ & $369 \sim 391 $& 0 & $-0.33 \sim -0.18$ & $0.28 \sim 0.79$ \\
& & $448 \sim 475 $& & $-0.26 \sim -0.14$ &$0.19 \sim 0.54$\\
\hline
\end{tabular}
\end{center}\label{results1}
\end{table}

 Now we have finished the  calculation of the moments
 $\langle \xi^n_{s(\sigma)} \rangle$ of twist-3 distribution
amplitudes for scalar mesons $a_0$, $K^*_0$ and $f_0$ in the
framework of QCD sum rules. With the results of $\langle
\xi^n_{s(\sigma)} \rangle$, it is straightforward to derive the
Gegenbauer moments $a_m$ and $b_m$ in Eq.~(\ref{an},\ref{bn}) using
Eq.~(\ref{gege}). The results for the first non-zero Gegenbauer
moments at 1GeV and 2.1GeV scales are shown in table~\ref{results1}
and \ref{results2}. They can be applied to various approaches
involving light cone distribution amplitudes of hadrons, such as
perturbative QCD approach \cite{PQCD}, QCD factorization approach
\cite{QCDF} and light-cone sum rules \cite{LCSR} etc. As mentioned
above, the odd moments of twist-3 distribution amplitudes for scalar
mesons $a_0$ and $f_0$ are zero due to conservation of charge parity
and flavor symmetry as explained in the introduction. As a
byproduct, we also collect the masses and decay constants of scalar
mesons in table~\ref{results1} and \ref{results2}. These masses
indicate that the ground state of $\bar q q$ scalars are probably
$a_0(1450)$, $K^*_0 (1430)$ and $f_0 (1710)$.

 In ref.~\cite{yangmz},
the authors also studied the mass and decay constant of scalar meson
$K^{*}_0$. Their results are $m_{K^{*}_0}=(1410 \pm 49)$ MeV and
$f_{K^{*}_0}=(427 \pm 85)$ MeV, which are consistent with our
results within error bar.

\begin{table}[tb]
\caption{Masses, decay constants and Gegenbauer moments from the
tensor sum rules (\ref{SR3},\ref{SR4}) at the scale $\mu=1
{\rm{GeV}}$ and 2.1 GeV (shown in the second line of each meson)  }
\begin{center}
\begin{tabular}
{|c|c|c|c|c|c|} \hline\hline state & $m${(\rm{MeV})}&
$\bar{f}${(\rm{MeV})}& $b_1(\times 10^{-2})$ & $b_2$ & $b_4 $\\
\hline $a_0$&$1270 \sim 1390$ &$325 \sim 350$& 0 &
 $0 \sim 0.058$ & $0.070 \sim 0.20$  \\
 & &$395 \sim 425 $ & & $0 \sim 0.041$ & $0.045 \sim 0.13$\\
\hline$K^*_0$&$1440 \sim 1550$ &$349 \sim 375$&  $3.7 \sim 5.5$ & $0 \sim 0.15$ & ---  \\
 & & $424 \sim 456 $&$2.8 \sim 4.2$ & $0 \sim 0.11$ & \\
\hline$f_0$&$1620 \sim 1710$ &$381 \sim 426$& 0 & $-0.15 \sim -0.088$ & $0.044 \sim 0.16$\\
 & & $463 \sim 518 $& & $-0.11 \sim -0.062$ & $0.028 \sim 0.10$\\
\hline
\end{tabular}
\end{center} \label{results2}
\end{table}

\section{Summary}
\label{summ}

In this work, we have studied the masses, decay constants and
twist-3 distribution amplitudes of scalar mesons based on the
renormalization group improved QCD sum rules. It is shown that the
mass sum rules for scalar mesons are not very satisfied, since the
Borel windows are a bit narrow for all the three scalar mesons. Our
results for the scalar meson masses show that the physical states
$a_0(1450)$, $K^*_0 (1430)$ and $f_0 (1710)$ are preferred to be the
ground state of scalar mesons. The sum rules for decay constants of
these three scalar mesons are very stable in a much broader Borel
window. The second and forth scalar moments of $a_0$ can be obtained
with 30\% and 35\% uncertainties respectively, while both the second
and forth tensor moments of $a_0$ can be derived within 30\%
uncertainties. As for the $K^*_0$ meson case, the first and second
moments of scalar density twist-3 distribution amplitude
$\phi_{K^*_0}^{s}$ are obtained under 15\% and 20\% uncertainties
respectively. The uncertainties can be reduced to 10\% for both the
results of the first and second moments of tensor twist-3
distribution amplitude $\phi_{K^*_0}^{\sigma}$. For the case of
$f_0$ meson, the second moment for both of scalar twist-3
distribution amplitude $\phi_{f_0}^{s}$ and tensor twist-3
distribution amplitude $\phi_{f_0}^{\sigma}$ each has 25\%
uncertainties. Besides, the fourth moment for each of these two
distribution amplitudes could be obtained within 30\% uncertainties.
It is also worthwhile to emphasize that the correlation functions
are calculated to leading $\alpha_s$ power based on operator product
expansion in this work, which will bring some additional
uncertainties to mass, decay constants and Gegenbauer coefficients.

It is found that the second Gegenbauer coefficients of scalar
density twist-3 distribution amplitudes for $K^{*}_0$ and $f_0$
mesons are quite close to that for $a_0$, which indicates that the
SU(3) symmetry breaking effect is tiny here. However, this effect
could not be neglected for the forth Gegenbauer coefficients of
scalar twist-3 distribution amplitudes between  $a_0$ and $f_0$.
Furthermore, one can also observe that the first two Gegenbauer
coefficients corresponding to tensor current twist-3 distribution
amplitudes for all the $a_0$, $K^{*}_0$ and $f_0$ are very small. As
is well known, the light-cone distribution amplitudes play a
critical role for hadronic decay processes in the framework of
factorization theorem where it describes the bound state effect of
hadrons. The available twist-3 distribution amplitudes of scalar
mesons allow us to improve the accuracy of the theoretical
predictions on the properties of scalar mesons, in particular for
the heavy flavor hadron decays to scalar mesons; so that it   is
very helpful for us to understand the structure of scalar mesons and
strong interactions.

\section*{Acknowledgements}

This work is partly supported by National Science Foundation of
China under Grant No.10475085 and 10625525. The authors would like
to thank H.Y. Cheng, T. Huang, X.H. Wu, K.C. Yang, M.Z. Yang, Y. Li,
Y.L. Shen and W. Wang for helpful discussions. C.D. L\"{u} thank
DESY theory group and Hamburg University for the hospitality during
his visit at Hamburg.

\end{document}